\RequirePackage{xcolor}

\documentclass[conference]{IEEEtran}
\IEEEoverridecommandlockouts
\usepackage{cite}
\usepackage{amsmath,amssymb,amsfonts}
\usepackage{algorithm}
\usepackage{listings}
\usepackage{graphicx}
\usepackage{textcomp}
\usepackage{subcaption}
\usepackage{hyperref}
\usepackage{cleveref}
\usepackage{xcolor}
\usepackage{enumitem}
\usepackage{amsmath}
\usepackage{comment}
\def\BibTeX{{\rm B\kern-.05em{\sc i\kern-.025em b}\kern-.08em
    T\kern-.1667em\lower.7ex\hbox{E}\kern-.125emX}}
\begin{document}

\title{Workrs: Fault Tolerant Horizontal Computation Offloading}

\author{
\IEEEauthorblockN{Alexander Droob, 
Daniel Morratz,}
\IEEEauthorblockN{Frederik Langkilde Jakobsen, \\
Jacob Carstensen, Magnus Mathiesen, \\
Rune Bohnstedt, and Michele Albano}
\IEEEauthorblockA{Department of Computer Science,
Aalborg University\\
Selma Lagerløfs Vej 300, 9220 Aalborg, Denmark\\
}
\and
\IEEEauthorblockN{Sergio Moreschini}
\IEEEauthorblockA{\textit{Tampere University} \\
\textit{Tampere, Finland}\\
Email: sergio.moreschini@tuni.fi}
\and
\IEEEauthorblockN{Davide Taibi}%
\IEEEauthorblockA{\textit{University of Oulu Oulu. Finland} \\
\textit{Tampere University. Tampere, Finland}\\
Email: davide.taibi@oulu.fi}
}

\maketitle

\begin{abstract}
The broad development and usage of edge devices has highlighted the importance of creating resilient and computationally advanced environments. When working with edge devices these desiderata are usually achieved through replication and offloading. 
This paper reports on the design and implementation of Workrs, a fault tolerant service that enables the offloading of jobs from devices with limited computational power. We propose a solution that allows users to upload jobs through a web service, which will be executed on edge nodes within the system. The solution is designed to be fault tolerant and scalable, with no single point of failure as well as the ability to accommodate growth, if the service is expanded. The use of Docker checkpointing on the worker machines ensures that jobs can be resumed in the event of a fault. 
We provide a mathematical approach to optimize the number of checkpoints that are created along a computation, given that we can forecast the time needed to execute a job. 
We present experiments that indicate in which scenarios checkpointing benefits job execution. 
The results achieved are based on a working prototype which shows clear benefits of using checkpointing and restore when the completion jobs' time rises compared with the forecast fault rate.
The code of Workrs is released as open source, and it is available at \url{https://github.com/orgs/P7-workrs/repositories}. 
This paper is an extended version of \cite{edge2023paper}.
\end{abstract}


\section{Introduction} \label{Chapter:Introduction}
In the past decade, more and more devices such as kitchen appliances, cars and even toilets have become part of what is called the Internet of Things (IoT). The Internet Of Things describes a network of physical objects connected to the internet, which  makes it possible to collect and exchange data with other systems and devices~\cite{IOTDefined}.

In 2022 more than 14 billion devices were connected to the IoT and this number is expected to reach over 27 billion by 
2025~\cite{IOTNumberOfDevices}. 
In fact, as time goes by, there is more access to  low-cost, low-power sensors making IoT technology more available. 
Even with this advance in technology, IoT still comes with some challenges. Many IoT devices have limited hardware capabilities such as processing power and memory, and some devices are lacking some components like local storage or Graphical Processing Units, moreover  
some IoT devices have custom hardware, which makes  application development tedious and can limit the availability of ready-to-use software libraries. 

To overcome the challenges posed by the lack of computational power and missing hardware, a widespread solution is to outsource computational jobs 
to other more powerful or more specialized machine. This concept is known as computation offloading~%
\cite{CompOffloading}, and its benefits comprise  
improved application performance and energy saving for  battery-driven devices. 


We consider that the platform targeted by the computation offloading process can either prioritize cost over reliability (in the case of edge nodes) or be not under the full control of the user in  the case of computation power provided by volunteers. To add resilience to faults that can interrupt the current job, it can be useful to create a checkpoint of the current status of the computation, move it to another device, and use it to continue the computation from the checkpoint in case a fault occurs. 

This work aims at creating an efficient and robust offloading solution based on edge nodes, by taking into account possible sources of faults:
\begin{enumerate}
    \item The worker running on an edge node can suffer a fault 
\item The  orchestrator that allocate jobs on  workers and keeps track of the checkpoints, can suffer faults itself
\item To identify the location (IP address, port, etc) of the components of the architecture, it is either necessary to have a register (e.g.: a  Dynamic DNS), or have all architectural components act like clients towards a message broker. Both cases represent the introduction of one more architectural component that can suffer a fault
\end{enumerate}

In our proposed solution, called Workrs, we adopt the utilization of a RabbitMQ message broker~\cite{dossot2014rabbitmq} in a clustered configuration~\cite{RabbitMQClustering} to take care of source of fault 3 from the list just presented. 
We replicate the orchestrator to protect our architecture against their faults, and use a totally ordered multicast, enabled by the clustered RabbitMQ Broker, to perform passive replication between them, to solve source of fault 2. 
Each worker periodically creates a checkpoint, and it sends  the checkpoint's  information (data dump, timestamp, job name, worker name) to an orchestrator  to solve source of fault 1. 

This paper presents the Workrs solution, which is robust by construction, and it analyzes it in terms of its efficiency. We study its performance in terms of total computation time for the submitted jobs, and energy expenditure. We consider that the baseline is a job run on a single worker without making use of checkpoints. If the worker fails frequently with respect to the length of the job, the total computation time can grow dramatically. Thus, making use of checkpoints can speed up the execution of a job. Regarding energy, consumption is proportional to the total computation time, plus an overhead related to the checkpointing procedure (which is not necessary if the architecture does not consider checkpointing). 
%




The rest of this paper is structured as follows:
Section~\ref{Chapter:ProblemDefinition} provides background information on the concepts and technologies employed in this work, and discusses related work;
Section~\ref{ssec:requirements} presents a MoSCoW analysis of the requirements considered for this work;
Section~\ref{sec:Proposed_Solution} describes the solution we created;
Section~\ref{Chapter:Experiments} reports on experimental results corroborating our approach; 
Section~\ref{Chapter:Conclusion} draws conclusions on the topic at hand and proposes future work.


\section{Background Information} \label{Chapter:ProblemDefinition}

This section  presents and analyzes some of the fundamental ideas behind computation offloading, and explores 
the technologies that will be used in this work. 

\subsection{Fault tolerant computation offloading}\label{SubSec:ComputationalOffloading}
A  computation offloading solution is inherently a  distributed system where components interact with each other by passing messages~\cite{offloading}. 
In fact, even though there are  benefits to computation offloading, 
it introduces multiple factors that have to be taken into account such as scalability, fault tolerance~\cite{fault1}, Quality of Service, deadlocks, communication capacity~\cite{IssuesDS} and 
robustness~\cite{CompOffloadingDesignFactors}. Moreover, introducing communication between devices over the internet can lead to high latency, if the required internet connection is slow or when transferring large amounts of data. 
Finally, security is another challenge that requires attention, since information is sent to a service outside of the user's control~\cite{CompOffloading}.

Computation offloading can either be vertical, for example from a mobile device to the cloud, or horizontal, with the computational job being sent from an edge node to another one. This work focuses on horizontal computation offloading, which is especially beneficial in cases where high latency can have a critical result in the performance of edge nodes processes~\cite{EdgeComputationVisionChallenges}. 



%

One recurrent issue in computing is making the system fault tolerant, meaning that the system can keep running as intended  in the occasion of partial failure~\cite{faultToleranceInDs}. In particular, fault tolerance comprises the properties of 
 %
    \emph{Reliability}, meaning that the system keeps  delivering what it is supposed to even when one or more components break down, and  
    \emph{Availability}, which involves that the system is ready to provide its functionalities even when a fault occurs. 

One of the most popular methods for dealing with fault tolerance is replication, involving creating one or more copies (replicas) of the system, and keeping them ready to take over  if the original system fails.

Another way to provide fault tolerance is checkpointing and recovery~\cite{check1}. This involves to create a snapshot of a running application, save it and then use the saved snapshot to recover the application and continue it from that point, should a fault occur. If checkpoints are taken regularly, checkpointing and recovery can be used to minimize the time loss in the event of faults.

\subsection{Technologies}
The goal of this section is to provide a concise introduction to the technologies and software used in this work.

\subsubsection{Docker}\label{dockerDesign}
To standardise the job uploaded from the clients to the workers, we will utilise the Docker framework~\cite{DockerContainer}. This framework allows to deploy a container image that wraps up code and dependencies and isolates the application thus making it possible to run the application reliably as a self-contained unit. 


An interesting feature of Docker can be leveraged to create a fault tolerant solution that can create checkpoints of the jobs. In fact, there is an experimental feature of Docker that makes use of an external tool  called Checkpoint/Restore In Userspace (CRIU)~\cite{DockerCheckpoint}. It works by freezing a running container and saving the state as a collection of files to disk thus making it possible to restore the container from the point it was frozen at. 

\subsubsection{Storage of checkpoints' information}
Based on the requirements that a user should be able to upload jobs, see the status of a job in progress and download the result, the orchestrator needs to be able to manage and store data. 
We decided to make use of a 
SQLite database~\cite{SQLiteBenefits},
which allows to embed the database itself in the host application. SQLite appears to be  faster compared to both a traditional database, and to applications that read and writ files using ad hoc functionalities~
\cite{SQLiteBenefits}. 
We aim at having a SQLite databases on the orchestrators, and having them replicated over the multiple replicas of these architectural components. 

\subsubsection{Message broker}

To implement communication between the components of our architecture, we decided to make use of a message broker, since it provides a number of advantages~\cite{ALBANO2015133}. 
RabbitMQ is an open-source message broker software that uses a producer/consumer model allowing producers to send messages to consumers. RabbitMQ brokers work by having producers send messages through exchanges that are responsible for routing these messages to various queues~\cite{dossot2014rabbitmq}. The queues act as pools where messages are stored. RabbitMQ uses acknowledgement-based message retention meaning that messages are retained by the queues and are only removed when the message has been sent to a consumer and the broker has received acknowledgement from the consumer~
\cite{RabbitMQ}. 
RabbitMQ brokers can also be set up in a cluster configuration, where
 a set of RabbitMQ instances   share users, exchanges, queues, runtime parameters etc. 
 This configuration  ensures failovers in case a RabbitMQ node becomes unavailable~\cite{RabbitMQClustering}.

\subsection{Related Work}

In a continuum environment characterized by devices with different nature and requirements, computation offloading plays a crucial role in ensuring efficient and effective use of resources~\cite{continuum}. Multiple works in the last decade have been showing the importance of computation offloading in such environment. 

In~\cite{COFF} Lin et al, performed a survey analyzing the first evolution of the trend of computation offloading towards edge computing. The work provided an overview of the different architectures as well as reviewed related works focused on  different key characteristics such as application partitioning, task allocation and resource management.

Mach et al. in~\cite{Mach} provided a different take on the subject, and  centered on user-centric use cases in mobile edge computing. 
Their study examined computation offloading decision, allocation of computation resources, and mobility management, comparing different works on these subjects.

Mao et al. proposed a dynamic computation offloading posing particular attention in energy harvesting technologies~\cite{Mao}. The technique  developed in the work supported  dynamic computation offloading by means of a   Lyapunov optimization-based dynamic computation offloading (LODCO) algorithm, which is characterized by low complexity, no prior information required, and takes into account the battery limitations of the devices. 

For what concerns specifically horizontal computation offloading, a two-step distributed horizontal architecture for computation offloading was presented in~\cite{deb}. In the work the horizontal offloading is mostly performed in the fog through directed acyclic task graphs.  
This results in an optimization of resources at the price of communication latency, which was justified in heavy computationally-required tasks. 

A solution based on checkpointing for system preservation was presented by Karhula et al. in~\cite{check2}. In this work, the checkpoint has been used to suspend long-running functions allowing Function as a Service and Serverless applications. Specifically, the checkpoint and restore approach has been used together with Docker to preserve the function states of the system enabling long-running functions. 

However, compared to all previously described works, together with taking into account the resource requirements for edge devices, we also aim at adding one more level of fault tolerance in the architecture by performing replication of the orchestrator and the message broker as well. Such replication ensures that our whole solution is fault tolerant. Moreover, we provide a mathematical formula to compute an optimal checkpointing strategy, and we corroborate our approach by means of experimental evaluation focused on both jobs execution time of computational jobs and energy consumption. 



\section{Requirements}\label{ssec:requirements}

Given the problems introduced in Section-\ref{Chapter:Introduction}, we define the Workrs solution in terms of its MoSCoW requirements~\cite{Moscow}. The list of requirements is as follows:


\begin{enumerate}[labelindent=1pt,label={[R}{\arabic*]}]
    \item A user must be able to upload a job to the solution.
    \item A user must be able to download the result.
    \item A user must be able to cancel a current job in progress.
    \item A user must be able to see the current status of a job.
    \item A user must be able to see previously completed jobs.
    \item A user must be able to see previously canceled jobs.
    \item A user must be able to identify themselves by providing a username
    \item The application should be able to distinguish between users.
    \item The application must be fault tolerant to the point where there is no single point of failure.
    \item The application must be scalable.
    \item A job must be resumable from a checkpoint
    \item A user must be able to see a graph visualization of the orchestrators and workers working on their jobs
    \item A user must be able to see statistics and performance information on their previous jobs
    \item The service could be secure
\end{enumerate}


Following the MoSCoW analysis, requirements \textbf{R1}, \textbf{R2}, \textbf{R7}, \textbf{R8} and \textbf{R11} have been identified as {\em must have}. Requirements \textbf{R9} and \textbf{R10} are {\em should have}.
Finally, requirements \textbf{R3}, \textbf{R4}, 
\textbf{R5}, \textbf{R6}, 
\textbf{R11} and \textbf{R12}  are {\em could have}.
Since our proposed solution will be implemented as a prototype only, security is not deemed as a priority. Given that security is a complex topic and can require a lot of effort on the design of a solution, requirement
\textbf{R14} was put into the {won't have} category. 


\section{The Workrs Solution}\label{sec:Proposed_Solution}

This section describes the proposed solution to the issues and requirements defined so far. 

\subsection{System Architecture}

\begin{figure}[t]
    \centering
    \includegraphics[width=0.48\textwidth]{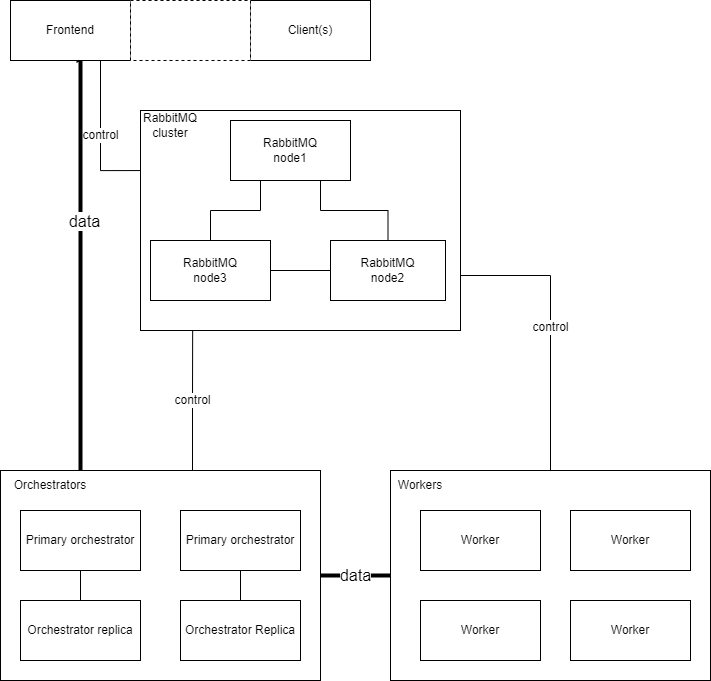}
    \caption{Overview of proposed architecture}
    \label{fig:proposedDesign}
\end{figure}

The architecture of our proposed computation offloading solution is illustrated in~\autoref{fig:proposedDesign}. 
The overall solution was designed with scalability and reliability in mind. In fact, the solution is designed so that the amount of orchestrators, clients and workers can be scaled up, and all architectural components are replicated so that there is no single point of failure.

The {\bf client} represents the computer utilized by the final user of the solution. 
The client will either communicate with a server-rendered webserver that will act as frontend, or it will itself run the frontend in a single page web application\cite{scott2015spa}. 


The {\bf frontend} is stateless and it provides a way for the client to interact with the rest of the solution. The frontend communicates with the orchestrators through the RabbitMQ cluster for control messages, and it can interact directly with the FTP server of an orchestrator for data messages, i.e.: to submit the script for job and to download the job results. 
Based on the requirements in \autoref{ssec:requirements}, the frontend exposes both an API (server-rendered web page), and a graphical user interface (single page web application) to provide the following functionalities to a user: 
\begin{itemize}
    \item Identification with a username
    \item Retrieve information about jobs uploaded by the user 
    \item Upload a new job to the solution 
    \item Download the result of a completed job
    \item Cancel a job in progress
    \item Graph visualisation of components in use
\end{itemize}

The {\bf orchestrator} is responsible for handling business logic in the solution between the frontend and workers, which connect  through RabbitMQ and using the FTP servers exposed by each orchestrator. 
When requested, the orchestrator will lookup relevant data on the client/worker association in a distributed hash table and respond to the client/worker with their ID in the orchestrator, which orchestrator will serve them will be responsible for their data. This allows the client or worker to initiate a session. 
When the orchestrator receives a job from a client via FTP the orchestrator creates makes relevant available to an available worker as a job, which will be notified through the RabbitMQ broker and will retrieve related data from the orchestrator's FTP server. 
When the orchestrator receives a checkpoint via FTP, it will be saved and in the event that a worker suffer from a fault, the job will be resumed from the latest checkpoint. When the orchestrator receives the result of a job from a worker via FTP, it will notify the  client who owns the job via the RabbitMQ broker to provide it with a  FTP download link for the result.

\begin{figure}[th]
    \centering 
    \includegraphics[width=0.48\textwidth]{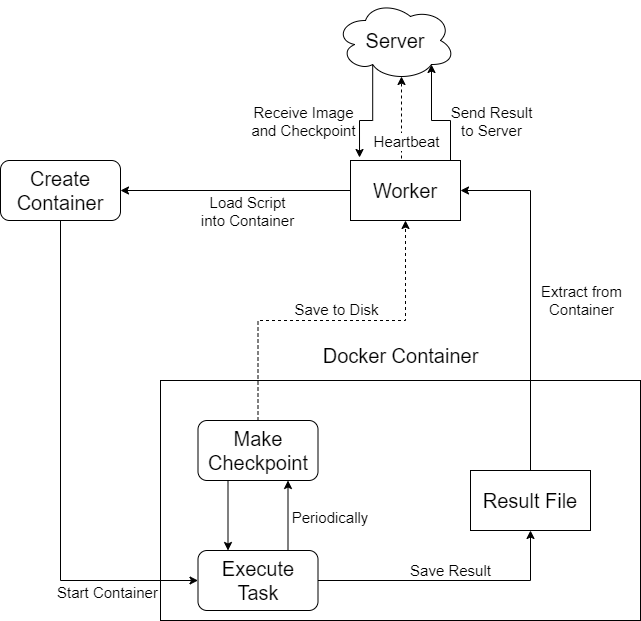}
    \caption{Flow diagram of a worker}
    \label{fig:workerFlow}
\end{figure}

The {\bf workers} are responsible for executing the jobs, checkpointing and resuming clients jobs. When a worker receives a job, the worker starts executing it. During execution the worker will periodically create checkpoints and upload  them to the orchestrator. The worker will also periodically send a heartbeat to the orchestrator to let it know that it is still operational. When a worker has finished executing a job, the result will be uploaded to the orchestrator and the worker will once again be ready to receive a new job. 
The execution flow of worker can be seen on \autoref{fig:workerFlow}.

The {\bf RabbitMQ cluster} is the message broker of the solution. 
The decision to use RabbitMQ for connecting various devices was made due to its ability to provide Quality of Service, which guarantees the delivery of sent messages between the devices. The RabbitMQ broker provides different options for how to communicate through the solution and for this service is decided to use a combination of queues and exchanges in a topic\cite{RabbitMQExchangesTopic}, direct\cite{RabbitMQExchangesDirect} and fanout\cite{RabbitMQExchangesTopic} configuration. By introducing a centralised broker the amount of connections within the solution is kept to only increase linearly when  edge node are joining the service thus making it scalable but creating other issues such as introducing the possibility of bottleneck and single points of failure. To overcome these issues RabbitMQ can be configured to run as a cluster increasing the throughput and having a failover strategy in case a RabbitMQ node becomes unavailable \cite{RabbitMQClustering}.

\subsection{Replication strategy}
Among the desired requirements, an essential {\em must have} was related to the possibility of resuming a job from a checkpoint. Resuming from a checkpoint results in a solution capable of creating snapshots of a running job periodically. The creation of snapshots would enable the chance of resuming a job from the latest snapshot whenever the edge node executing the job suffers from a fault. 
We decided to use the  Checkpoint/Restore In Userspace (CRIU)~\cite{DockerCheckpoint} of Docker to this aim.

With regards to the orchestrator, for sake of ensuring reliability and availability, we decided to have backup orchestrators that can take over for the primary orchestrator if this latter  component experiences hardware or software failure. 
We decided to implement this by means of passive replication, which is implemented by having a primary replica manager and one or more backup replica managers that can act as the primary in case of a replica fault. With regards to control messages, the primary replica will receive requests from a frontend, relay all requests to the other replica, and acknowledge the requests to the fronted only after the replicas confirm them. With regards to data, i.e. the scripts to be run as jobs and the results from the workers, the primary orchestrators saves them into a folder shared with the orchestrator replica, to let the operating system perform the replication. 

\subsection{Consistency strategy}
To maintain consistency across the primary and the backup orchestrators when they receive novel information such as the presence of a new checkpoint, it is important to make sure the all the orchestrator replicas would reach the same state when targeted by requests. It is therefore necessary to notify the orchestrator on changes of a shared resource, and to enxure that all the changes are applied in the same order.  
It is therefore important to use a multicasting strategy that provides total ordering, which is a communication procedure where a message is sent to a set of receivers, with all messages being received in the same order. 

\begin{figure}[t]
    \centering 
    \includegraphics[width=0.48\textwidth]{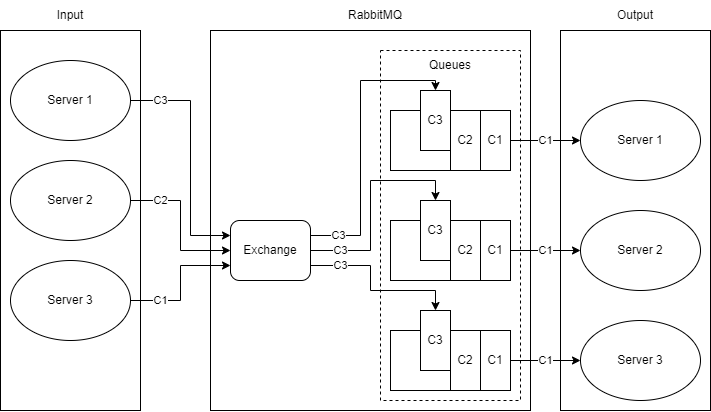}
    \caption{Example of multicast communication using RabbitMQ fanout exchanges}
    \label{fig:multicastStrategy}
\end{figure}

RabbitMQ provides fanout exchanges which allows for messages targeting the exchange to be received by multiple queues in the order in which the messages were received by the exchange. This means that RabbitMQ fanout provides multicasting with total ordering. Thus RabbitMQ fan out fulfills the solution needs for multicasting. 
An example of the chosen multicast strategy is illustrated in \autoref{fig:multicastStrategy} where three orchestrators push messages C1, C2 and C3 at the input side. The three messages arrive at the RabbitMQ exchange in the following order C1, C2 and C3 which is now replicated to all queues and are consumed by each orchestrator on the output side in the order the messages arrived to the exchange.



Here the totally ordered multicast using RabbitMQ as described above is taken advantage of. When an orchestrator wishes to make a change to a shared resource it will notify all known orchestrators and each one will then stop accessing the shared resources and respond that they are ready to receive a change. The orchestrator wishing to make the change will then publish the change to the fan out meaning that all orchestrators, including itself, will receive the change and once all known changes are consumed from an orchestrator queue and has been applied, the orchestrator will release the lock.

\begin{figure}[t]
    \centering 
    \includegraphics[width=0.48\textwidth]{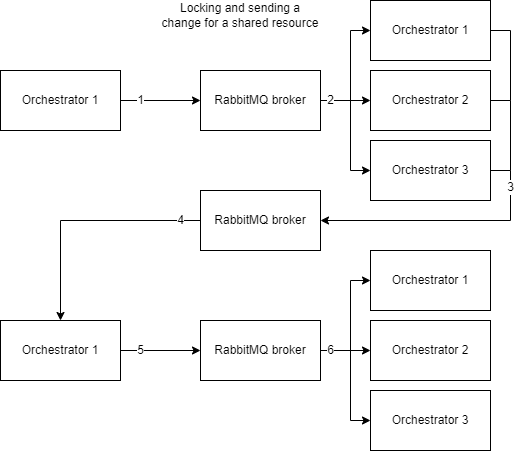}
    \caption{Illustration of a change procedure of a shared resource}
    \label{fig:lockRequest}
\end{figure}

\autoref{fig:lockRequest} shows  how an orchestrator initiates the locking procedure for a change of the shared resource:

\begin{enumerate}
    \item A locking request including a Globally Unique Identifier (GUID) is sent to the RabbitMQ multicast.
    \item The RabbitMQ broker takes the request and multicasts the request to all orchestrators.
    \item When each orchestrator reaches a safe state and it is ready to update the shared resource, it  sends an accept locking acknowledgement to RabbitMQ targeting the orchestrator requesting to lock the shared resource. 
    \item The RabbitMQ broker forwards all the replies to the requesting orchestrator. 
    \item When the requesting orchestrator has received acknowledgement from all orchestrators, it publishes the change to the RabbitMQ multicast along with the GUID associated with the change.
    \item RabbitMQ multicast the change to all orchestrators. When an orchestrator sees the change it validates the change by comparing the included GUID with the GUID from step 1 and performs the change on the locally stored shared resource. It then checks if further changes have been requested, otherwise it unlocks the shared resource and resumes normal operation.
\end{enumerate}

\subsection{RabbitMQ configuration}

\begin{figure}[th]
    \centering 
    \includegraphics[width=0.48\textwidth]{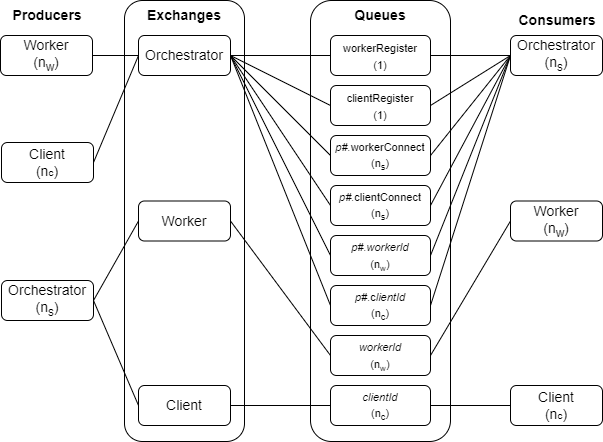}
    \caption{Overview of exchanges, queues and consumers where $n_s$ is the number of orchestrators in the solution, $n_w$  is the number of workers, $p\#$ is the orchestrator number and $n_c$  is the number of clients}
    \label{fig:RabbitMQOverview}
\end{figure}

The solution uses a RabbitMQ broker for communication, whose configuration is reported on \autoref{fig:RabbitMQOverview}. 
RabbitMQ topic exchanges leverage routing keys to direct messages to their appropriate queues, with each component of the architecture consuming messages from its designated queue. The broker contains three  exchanges, and all messages targeting the orchestrators are routed through the orchestrator exchange, messages targeting clients through the client exchange and messages targeting workers through the worker exchange. 


\begin{figure*}[t]
    \begin{subfigure}{0.32\textwidth}
        
\includegraphics[width=\textwidth]{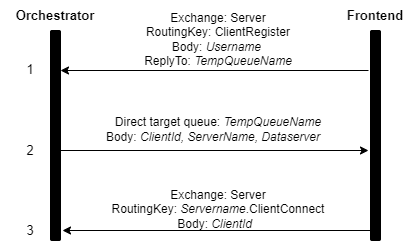}
    \caption{Overview of the client connection flow}
    \label{fig:ClientConnectionDiagram}

    \end{subfigure}
\hfill
    \begin{subfigure}{0.32\textwidth}
\includegraphics[width=\textwidth]{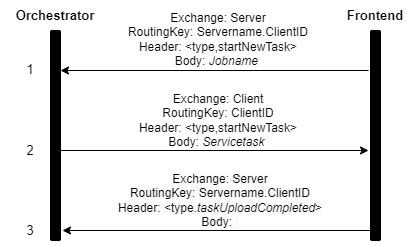}
    \caption{Overview of the job uploading flow}
    \label{fig:UploadJobDiagram}
    \end{subfigure}
\hfill
    \begin{subfigure}{0.32\textwidth}
\includegraphics[width=\textwidth]{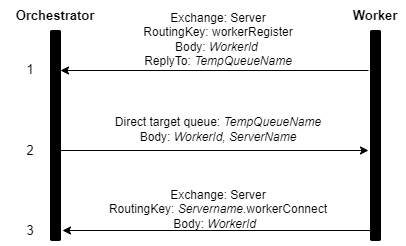}
    \caption{Overview of the worker connection flow}
    \label{fig:WorkerConnectionDiagram}
\end{subfigure}
\caption{Examples of communication between frontend, orchestrator and worker}

\end{figure*}
\paragraph{Client connection flow}
%
\autoref{fig:ClientConnectionDiagram} shows the communication flow of the client that connects to the solution through its frontend. First the client registers for a session by sending a username to the orchestrator exchange with routing key "clientRegister". Along with the message, the client sends the name of a temporary queue that was created by the client to receive a response. 

After this the orchestrator that happened to consume the session registration from the client will look up client information using the received username. The orchestrator responds to the client by directly targeting the temporary response queue. The response contains a client ID and the name of the orchestrator that will serve the client. 
The client saves the received information, discards the temporary queue,  creates a client queue  and binds it to the client exchange with routingkey "\{clientId\}", and adds a consumer to this queue that will receive all future messages for the client.

Lastly the client sends a connection request to the orchestrator that the client was told will serve it by targeting the orchestrator exchange with routing key "\{orchestratorname\}.clientConnect". Here the client provides its client id. Upon receiving the message the orchestrator will set up a consumer on the queue bound to the orchestrator exchange with routing key "\{orchestratorname\}.\{clientId\}", to receive future messages from the client. 

\paragraph{New job flow}
\autoref{fig:ClientConnectionDiagram} shows the communication flow when a client communicates with an orchestrator requesting to create a new job. The client sends a message to the orchestrator exchange with the routing key "\{orchestratorname\}.\{clientId\}", the header "$<$type,startNewTask$>$" and the job name. 

The orchestrator then responds acknowledging  the newly created job, comprising a link to a folder in the  FTP server running on the selected orchestrator. The client then uploads the script for the job on the orchestrator's FTP server, and then sends  a  "$<$type.taskUploadCompleted$>$" message to the orchestrator. 

\paragraph{Worker connection flow}
The communication flow of a worker connecting to the service can be seen on \autoref{fig:WorkerConnectionDiagram}. 

The first message from worker to orchestrator provides a worker id rather than a username when registering for a session, and creates a temporary queue to receive a response. 
The very first time a worker connects it will provide an empty workerId and the orchestrator will create a new id for the new worker. If the worker has connected before it will have already saved the id on its  edge node and will provide this when registering. When responding to the session registration, the orchestrator will respond to the temporary queue with the created or provided worker id and the name of the orchestrator that should serve the worker and the worker creates a consumer on the queue bound to the worker exchange with routing key "\{workerId\}". 
Lastly,  the worker sends a connection request to the orchestrator that it will server it, and the orchestrator creates a consumer on the queue bound to the orchestrator exchange with routing key "\{orchestratorname\}.\{workerId\}".

\subsection{Optimal frequency of checkpointing}\label{ssec:optimalcheckpoints}

In case both the faults frequency and the total time to perform a computation can be forecast, it is possible to compute the optimal frequency for the checkpoints to minimize the total execution time of the job. 
This section will use the following definitions: 

\begin{itemize}
    \item $T$ = Total time to complete a job
    \item $\mu$ = Probability of fault in the unit time
    \item $p(t)$ = Probability density for a fault

    \item Execution time per part, with checkpointing
    \item $n$ = Time between checkpoints
    \item $overhead$ = Cost of checkpointing
\end{itemize}

We model the fault's distribution as a Poissonian:

\begin{equation}\nonumber
p(t) = \mu e^{-\mu t}
\end{equation}



The time to complete a job (see Appendix~\ref{app:calcoli}), given that  faults can occur and they  lead to restarting the job, is given by: 

\begin{equation}\nonumber
E_x(\mu, T) = \frac{e^{\mu T}-1}{\mu}
\end{equation}

When checkpointing is part of the picture, the process is essentially split into a set of $N \in \{1, ...\}$ processes of length $T/N$ by making use of $N-1$ checkpoints, at the cost of an overhead of size $C$ for each checkpoint, thus the total execution time becomes:

\begin{IEEEeqnarray}{l}\label{eq:optimalcost}
    N E_x\left(\mu,\frac{T}{N}\right) + (N-1)~C 
\end{IEEEeqnarray}

Equation~\ref{eq:optimalcost} is convex (see Appendix~\ref{app:calcoli}), thus it is possible to find the optimal number of checkpoints to be used by means the  algorithm reported in Listing~\ref{alg:optimal}).


\begin{lstlisting}[language={[Sharp]C}, caption={Optimal number of checkpoints, given that $T$ and $C$ can be forecast}, label={alg:optimal}]
int optimal_checkpoints_numer(mu, T, C):
    bool in_progress = true;
    int best_N = 0;
    double best_time = predict_time(mu, T);
    do {
        N = N + 1;
        double exec_time = N * 
            predict_time(mu, T / N) + (N-1) * C;
        if (exec_time > best_time) {
            in_progress = false;
            N --;
        } else {
            best_time = exec_time;
        }
    }
    return(N);
}

double predict_time(double mu, double T) {
    return (Math.Exp(mu * T) - 1) / mu;
}
\end{lstlisting}


\subsection{Frontend Implementation} \label{Chapter:Implementation}

We chose to implement the {\bf frontend}  in the .NET framework Blazor. Blazor allows for server-side rendering, making it possible to send events from the browser to the orchestrators through it~\cite{BlazorOverview}. Additionally, Blazor is component-based, written in C\# and it  can create interactive user interfaces. Moreover, the C\# libraries to interact with a RabbitMQ server proved to work "as they were", while other libraries proved to have issues such as incompatibilities with clustered RabbitMQ brokers.

\section{Experiments} \label{Chapter:Experiments}

This section describes the experiments performed over the solution, and their results.

\subsection{Experimental Deployment}

\begin{figure}[b]
    \centering
    \includegraphics[width=0.48\textwidth]{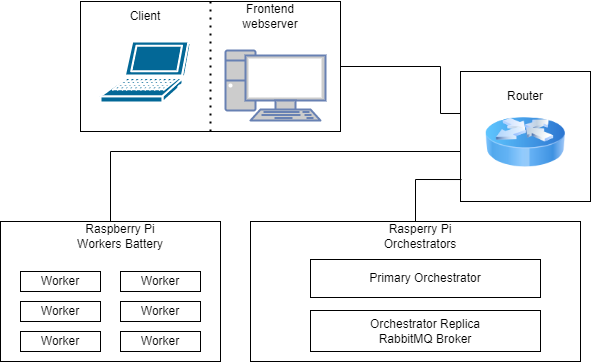}
    \caption{Logical representation  of the experimental setup}
    \label{fig:prototypeOverview}
\end{figure}

\begin{figure}[t]
    \centering
    \includegraphics[width=0.48\textwidth]{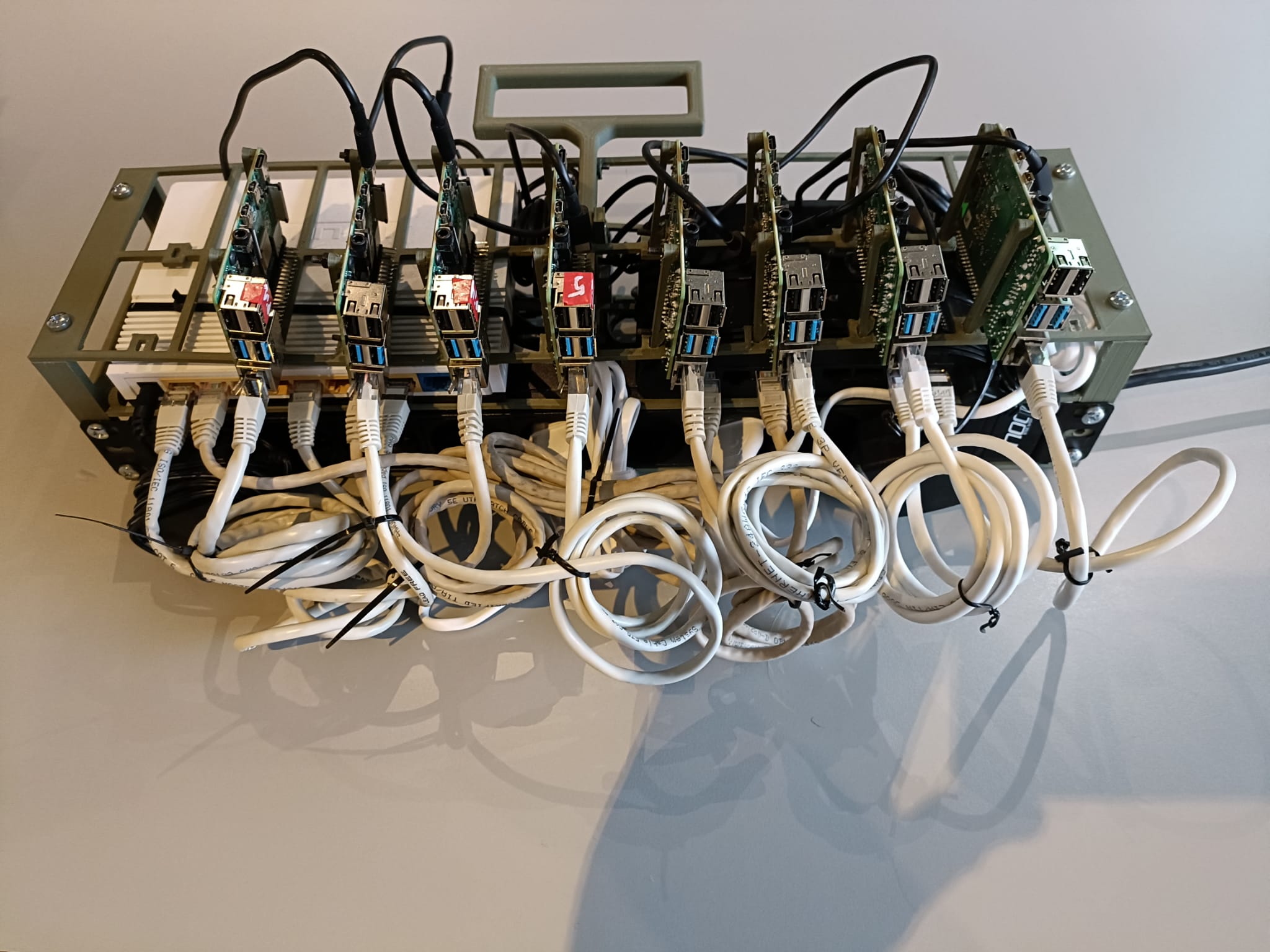}
    \caption{Raspberry pi cluster used in the experiments}
    \label{fig:raspicluster}
\end{figure}

The prototype we created runs on a Local Area Network (LAN) containing one personal computer and  8 edge nodes. A logical representation of this setup is illustrated on~\autoref{fig:prototypeOverview}, and the physical Raspberry pi cluster we built is shown in~\autoref{fig:raspicluster}.

The edge nodes are Raspberry Pis 4 with 1 GB Ram and 16 GB of removable SD storage for the main drive. The operating system flashed to the Raspberry Pi is Ubuntu Server 20.04.5 LTS (64-bit). The containers are run using Docker Engine 20.17. 

Six of the edge nodes run one worker each. One more edge node runs the primary orchestrator. The last edge node runs the orchestrator replica and the RabbitMQ broker. Since the experiments focus on corroborating the formulas from \autoref{ssec:optimalcheckpoints} and the fault tolerant orchestrator and workers deployment, we did not set up a clustered RabbitMQ broker, but from its specifics it appears that it would have not impacted the message bandwidth, nor it would have been part of the trade-off that we are evaluating~\cite{RabbitMQClustering}. 
The replication of the checkpoint files over the secondary orchestrator is performed by saving the files on a folder shared via the SMB protocol~\cite{bettany2015managing} between the orchestrators, thus it happens asynchronously with respect to the rest of the checkpointing functions and it does not impact the performance of the solution. 

We have power meters in place to measure the energy spent by each edge node. 
However, in the experimental results we will show only the total energy spent by  the two orchestrators and two workers, one executing the job and the other one ready to take over if any fault occurs, since the other edge nodes were not involved in the experiments.

A desktop computer (whose energy consumption we did not measure) runs the client and the frontend web server. 
%
%
Initally, the idea was to create a Docker image with the Python interpreter, its libraries and the script related to a job every time a job is submitted. However,  it was discovered that building the image in the orchestrator would introduce unnecessary overhead during job startup, since all images were identical except for the Python script to be executed. Thus, we created one base image containing the  Python interpreter and some of its most useful libraries, and we pre-installed it on all the edge nodes. The worker would then download  the Python script  for the particular job to be executed, it would  run the base image, and inject into it the Python script. The  job startup time got much smaller, since the download time for the Python script is much lower than downloading the full Docker image.

\subsection{Checkpoint time penalty} \label{sec:checkpointTimePenalty}

The first  question we aim to answer regards the overhead  incurred by periodically creating checkpoints of the jobs and uploading them to the primary orchestrator. To this aim, we ran a series of jobs, each of them having a completion time of $300$ seconds, and we set a very low $\mu$, meaning that we expected to have no faults during the job execution. 

The first set of jobs were executed without doing any checkpoint, then more jobs were executed performing a checkpoint every $18.75$ seconds of job execution (meaning that we stopped the checkpoint timer while performing the checkpoint itself), thus performing a total of $15$ checkpoints. 

\begin{figure}[t]
    \centering
    \includegraphics[width=0.48\textwidth]{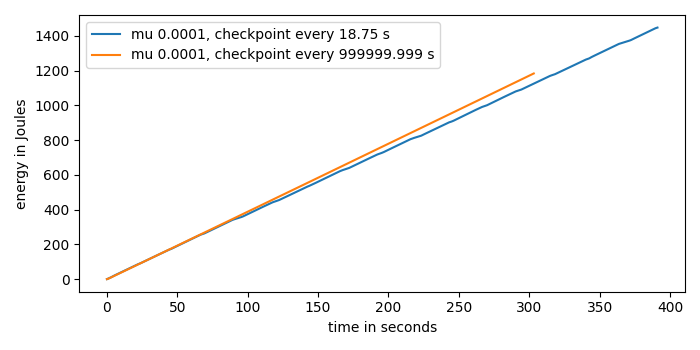}
    \caption{Execution time $300s$, no faults, comparison between no checkpoints and $15$ checkpoints.}
    \label{fig:CheckVSnoCheck}
\end{figure}

%
Figure~\ref{fig:CheckVSnoCheck} shows the time to complete the job on the $x$ axis, and the energy spent on the y axis. 
The results  hint that the cost of checkpointing is approximately $90$ seconds in total, i.e.: each checkpoint causes a delay of $6$ seconds. 
%
When computing Eq.~\ref{eq:optimalcost} in 
the rest of this section, we will consider the cost of checkpointing as $6$ seconds.

\subsection{Fault Time penalty} \label{sec:FailureTimePenalty}

\begin{figure}[b]
    \centering
    \includegraphics[width=0.48\textwidth]{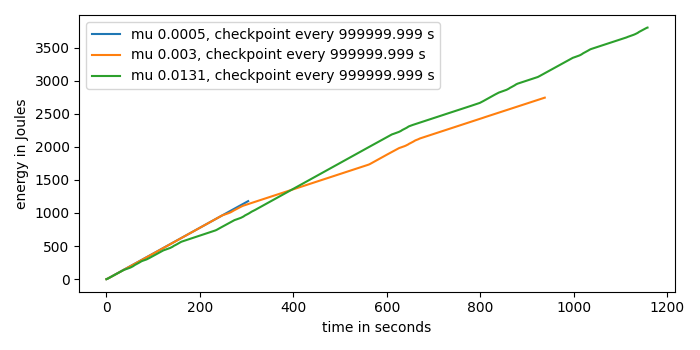}
    \caption{Execution time $300s$, different  faults frequencies, no checkpoints.}
    \label{fig:checkpointing_required}
\end{figure}

To assess if checkpointing is required given the experiments' parameters, we did not  perform any checkpointing with different $\mu$. Figure~\ref{fig:checkpointing_required} shows that a low $\mu$ is perfectly compatible with not using checkpointing, while a high $\mu = 0.131$ leads to a very long execution time. 

\begin{figure}[t]
    \centering
    \includegraphics[width=0.48\textwidth]{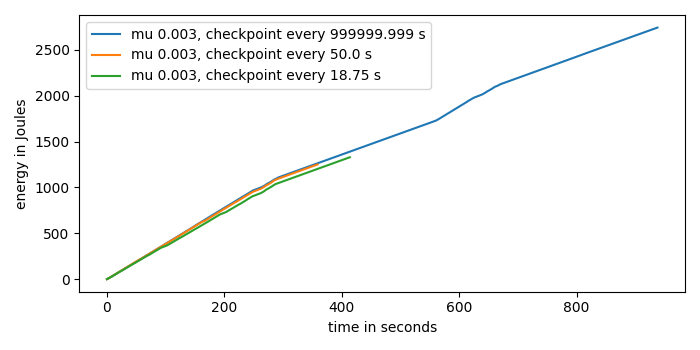}
    \caption{Execution time $300s$, frequent   faults, different  checkpointing frequencies.}
    \label{fig:best}
\end{figure}

To understand how often to perform a checkpoint, and to corroborate the formulas and algorithm from Section~\ref{ssec:optimalcheckpoints}, we set a relatively high $\mu = 0.003$. 
We performed experiments with no checkpoint, with $15$ checkpoints, and with $5$ checkpoints (one checkpoint every $50$ seconds), as suggested by the algorithm in Listing~\ref{alg:optimal}. Figure~\ref{fig:best} shows that the best results correspond to a checkpoint every $50s$, corroborating our formulas.

\section{Conclusion and Future Work}\label{Chapter:Conclusion}

The scope of this work was to create a robust distributed system for  computation offloading. By focusing on creating a solution with high fault tolerance 
(see the  MoSCoW analysis in \autoref{ssec:requirements}) 
every architectural component was designed with the goal of 
eliminating single points of failure while remaining scalable. For sake of focusing on the problem at hand, current prototype had to sacrifice on other characteristics such as system security. 
%

In \autoref{Chapter:Experiments} the cost-benefit of checkpointing was investigated and the expected total execution time for jobs of varying length both with and without checkpointing with different fault rates were compared. The results confirm experimentally that checkpointing is more useful when the jobs' completion time gets larger with respect to the expected fault rate. 
A limitation of the study is that in the real world  the fault rate might be hard to forecast and dependent  on the environmental conditions the workers experience, and the completion time for a job can be even harder to forecast. 

On the positive side, the experiments were performed considering that  both workers and orchestrators were run on the  same kind of edge device. In a real deployment, the orchestrator would be deployed in a more energy-saving device (e.g.: Raspberry pi), the node devices running workers would be more energy hungry (e.g.: Jetson Xavier), and the ratio worker edge nodes / orchestrator edge nodes would be much higher than in the current work.





\subsection{Future Work} \label{Chapter:FutureWork}
Future work was planned to  improve the solution on different aspects. 

\paragraph{API}
As this solution  is aimed towards users with a software background, it could be beneficial to provide an API interface, making it possible to gain programmable access to the solution. 

\paragraph{Frontend as a One Page App}
We are considering to create a Javascript One Page App that will integrate the frontend within the client, instead of a server-rendered web page, to interact with the RabbitMQ broker, thus allowing the client to be independent from a webserver. 

\paragraph{User interface}
The user interface is still a prototype, and its future implementation will comprise an interactive admin tool for administrators, and a  user interface that  provides useful settings  such as enabling or disabling the checkpointing feature or specifying the interval between checkpoints dynamically. 

\paragraph{Security}
If the solution  were to be publicly available, security measures should have been taken into account. This could include password-based authentication and validation of users. Furthermore, a user currently has no limitation in the system and could potentially perform a Denial of Service attack by overworking all edge nodes and thus interrupting the normal operation of the solution. 

\paragraph{Support for more languages}
Right now it is only possible to upload python scripts only  as jobs. Future versions could add support for more languages. The orchestrators could then create images specific to each language, to be used as a basis for the execution of the users' jobs. For example, a C\# image could be created with a pre-installed .Net Core environment as well as the ability to download a specified list of Nuget packages. Likewise, a JavaScript image could have a Node.JS, Dine, ExpressJS or similar environment pre-installed, and download NPM packages.


\section*{Acknowledgments}


This work  was partially supported by 
the S4OS Villum Investigator Grant (37819) from Villum Fonden, by Industry X and 6GSoft  projects funded by Business Finland.

\bibliographystyle{IEEEtran}

\bibliography{litteratur.bib}

\appendices

\section{Computation of Best Checkpoint Timing}\label{app:calcoli}

This section will calculate and analyse the total expected execution time of a script with and without checkpointing, and provide formal proof for the formulas and the algorithm in \autoref{ssec:optimalcheckpoints}. 
This section uses the same definitions of \autoref{ssec:optimalcheckpoints}.

We model the fault's distribution as a Poissonian:

\begin{equation}
p(t) = \mu e^{-\mu t}
\label{eq:pt}
\end{equation}

Equation~\ref{eq:pt} dictates that the greater \textit{t} is, the smaller value of $p(t)$, and in particular that, given that no fault happened before time $t_0$, the probability of a fault between $t_0 + t_1$ and $t_0 + t_2$ do not depend on $t_0$, i.e., $p(t)$ is a memoryless function. 

The probability for a fault to happen before time $T$ is:

\begin{equation}
P_F(\mu, T) = \int^T_0 p(t) \,dt = 1 - e^{-\mu T} \nonumber
\end{equation}

If we have the a priori knowledge that a fault occurs before time T, the expected time for the fault is:

\begin{IEEEeqnarray}{l}
T_F(\mu, T) = 
\frac{\int_0^T t p(t) \,dt}{\int^T_0 p(t) \,dt} = \nonumber \\
\frac{
\left[-t \cdot e^{-\mu t}  - \dfrac{e^{-\mu t}}{\mu}  \right]_{0}^{T}
}{
\left[ -e^{- \mu t} \right]_0^T
} = \nonumber \\
\frac{\frac{1}{\mu} - T e^{-\mu T} - \frac{e^{-\mu T}}{\mu}}{1 - e^{-\mu T}} \nonumber
\end{IEEEeqnarray}



For the sake of clarity, in the following formulas we will use $P_F = P_F(\mu, T)$ and $T_F = T_F(\mu, T)$.
The time to complete the task, given that faults lead to restarting the computation, can be computed as the probability of having $0$ faults multiplied by $T$, plus the probability of having $1$ fault multiplied by the sum of $T$ and $T_F$, plus the probability of having $2$ faults multiplied by $T + 2 T_F$, etc, leading to the expected execution time:

\begin{IEEEeqnarray}{l}\label{eq:execution_time}
E_X(\mu, T) =
(1 - P_F) T + (1 - P_F) P_F (T + T_F) + \nonumber \\ 
(1-P_F) P_F^2 (T + 2 T_F) + ... = \nonumber  \\
(1-P_F) \sum_{i=0}^\infty
P_F^i ( T + i T_F ) = \nonumber  \\
e^{-\mu T} ( T \frac{1}{1-P_F} + T_F \frac{P_F}{(1 - P_F)^2}) = \nonumber  \\
\frac{e^{\mu T}-1}{\mu} \nonumber
\end{IEEEeqnarray}


When checkpointing is part of the picture, the process is essentially split into a set of $N \in \{1, ...\}$ processes of length $T/N$, at the cost of an overhead of size $C$ for each of the $N-1$ checkpoints, thus the total execution time becomes:

\begin{IEEEeqnarray}{l}
E_Y(\mu, T, N, C) =
    N E_x\left(\mu,\frac{T}{N}\right) + (N-1)~C \label{eq:ey}
\end{IEEEeqnarray}

It is interesting to study how Eq.\ref{eq:ey} depends on $N$:

\begin{IEEEeqnarray}{l}
\frac{d\,E_Y(\mu, T, N, C)}{d\,N} = \nonumber \\
E_x\left(\mu,\frac{T}{N}\right) + N \frac{-e^{\mu T / N} \mu T}{\mu N^2} 
+ C = \nonumber \\
\frac{
N e^{\mu T / N} - N 
- \mu T e^{\mu T / N}
+ C \mu N
}{\mu N} = \nonumber \\
\frac{e^{\mu T / N} -1}{\mu} 
- \frac{T}{N} e^{\mu T / N}
+ C \nonumber 
\end{IEEEeqnarray}

The second derivative is calculated as:

\begin{IEEEeqnarray}{l}
\frac{d^2\,E_Y(\mu, T, N, C)}{d\,N^2} = \nonumber \\
-\frac{T}{N^2} e^{\mu T/N} + \frac{T}{N^2} e^{\mu T /N} + \frac{\mu T^2}{N^3} e^{\mu T /N} = \nonumber \\
\frac{\mu T^2}{N^3} e^{\mu T /N} \label{eq:deriv2}
\end{IEEEeqnarray}

Given that we are considering $\mu>0$, $T>0$ and $N\ge 0$, Eq.~\ref{eq:deriv2} is $\ge 0$, thus $E_Y(\mu, T, N, C)$ is convex and it has either one minimum for $N>0$, or an absolute minimum in $N = 0$. This proves that Algorithm~\ref{alg:optimal} computes the best checkpointing strategy correctly.

\end{document}